\documentclass[preprint]{elsarticle}

\usepackage{amsthm,amsfonts,amssymb,mathrsfs,showidx,arydshln}
\usepackage[fleqn]{amsmath}
\usepackage{lineno,hyperref,amsmath}
\modulolinenumbers[100]

\newcommand{\n}{\hat{n}}

\begin{document}

\begin{frontmatter}

\title{A $q$-deformation of the Bogoliubov transformations}

\newtheorem{sol}{Solution}
\newtheorem{exe}{Example}

\author[add1]{Ivan Arraut}
\ead{ivanarraut05@hotmail.com}
\author[add2]{Carlos Segovia}
\ead{csegovia@matem.unam.mx}

\address[add1]{Department of Physics, Faculty of Science, Tokyo University of Science,
Tokyo, Japan.}
\address[add2]{Instituto de Matem\'aticas UNAM, Oaxaca, M\'exico.}

\begin{abstract} 
An approach for $q$-deformed Bogoliubov transformations is presented. 
Assu\-ming a left-right module action together with an $*$-operation and 
 deformed commutation relations, we construct a $q$-deformation of the nonlinear Bogoliu\-bov transformation. 
 Finally, we introduce a Hopf structure when $q$ is a root of unity.

\end{abstract}

\begin{keyword} Bogoliubov transformation, Hopf algebras
\MSC[2010] 16T05
\end{keyword}

\end{frontmatter}

\linenumbers

\section{Introduction}

The Bogoliubov transformations defined in the theory of superconductivity \cite{Bog0, Val}, connect different notions of vacuum by defining different sets of annihilation and creation operators, 
furthermore, they are used as a method to analyze particle creation in quantum field theory \cite{Parker}.  Although there are analogues to the $q$-Bogoliubov transformation in \cite{Zhe} and \cite{Kat}, it is infeasible to construct a $q$-deformed Bogoliubov preserving the $q$-commutation relations \cite{Kat}. 
The theory of quantum groups \cite{Man} indicates how to construct the symmetries of quantum planes, where the coefficients are commuting sets. 
In contrast, in this work we assume deformed commutating sets and we obtain interesting identities for the coefficients of the $q$-Bogoliubov transformation, which is an analog of the work of Katriel \cite{Kat}.
In this work we promote the Bogoliubov coefficients to operators and we assume some deformed commutation relations, which make it possible to develop the $q$-deformed Bogoliubov transformations. Future application of these results have the purpose of constructing non-commutative models of Hawking radiations as has been done in \cite{Xin}. 
  \section{Bogoliubov transformations}
 Consider the annihilation operator $A$ and the creation operator $A^*$ together with the Bogoliubov coefficients operators $U_n$ and $V_n$ for $n\in\mathbb{N}\cup\{0\}$. The $*$-operation is an antihomomorphism given by sending $A$ to $A^*$ and $U_n,V_n$ to the adjoints and conjugation for constants. Consider unitary complex numbers $z$ and $w$  such that
\begin{equation}
zw=1\,.
\end{equation}
 Assuming $[A,A^*]_q=AA^*-qA^*A=1$ and the deformed commutations
\begin{equation*}
X_{n+1}A=zAX_n\textrm{ and }AX_n^*=wX_{n+1}^*A \,,
\end{equation*}
where $X_n\in \{U_n,V_n\}$. 
Introducing 
\begin{equation}   \label{Bogotrans}
T(A)=U_nA+V_{n-1}A^*\hspace{1cm}\textrm{and}\hspace{1cm}T(A^*)=A^*U_n^*+AV_{n-1}^*\,.
\end{equation}
For $Q$ a real number the $Q$-bracket has the form 
\begin{equation*}
[T(A),T(A)^*]_Q =w^2 \Phi(n)	A^2+z^2{A^*}^2\Phi(n)^*+w^2\Psi(n) \,,\end{equation*}
where $\Phi(n)=U_nV_{n+1}^*-QV_n^*U_{n+1}$ and
\begin{equation*}
\Psi(n)=U_nU_n^*AA^*-QU_{n-1}^*U_{n-1}A^*A+V_{n-1}V_{n-1}^*A^*A-QV_n^*V_nAA^*\,.
\end{equation*}
Thus for $\Phi(n)=0$ and $\Psi(n)=\frac{1}{w^2}$ we obtain the identity $[T(A),T(A)^*]_Q=1$. Assuming 
\begin{equation}
U_nU_n^*-QV_n^*V_n=\frac{1}{w^2}\textrm{ and }V_nV_n^*-QU_n^*U_n=-\frac{q}{w^2}\,.
\end{equation}
We obtain immediately $\Psi(n)=\frac{1}{w^2}$.

\section{Nonlinear $q$-Bogoliubov transformations}
\label{sec1}
The present section gives an analog of the work of Katriel \cite{Kat}. We consider $[A,A^*]_q=AA^*-qA^*A=1$ with the $q$-Fock states defined as usual
\begin{equation*}
|k\rangle_q=\frac{(A^*)^k}{\sqrt{[k]_q!}}|0\rangle_q\,,
\end{equation*}
where the $q$-vacuum state satisfies $A|0\rangle_q=0$, and where $[k]_q=\frac{q^k-1}{q-1}$, $[0]_q!=1$, and $[k]_q!=[k-1]_q![k]_q$. 
We have the relations
\begin{equation*}
A^* |k\rangle_q= \sqrt{[k+1]_q}|k+1\rangle_q\textrm{ and }
A|k\rangle_q = \sqrt{[k]_q}|k-1\rangle_q\,.
\end{equation*}
Thus $A^* A=[\n]_q=\frac{q^{\n}-1}{q-1}$, where $\n$ is the number operator $\n$ given by $\n|k\rangle_q=k|k\rangle_q$. Notice 
$AA^*=qA^*A+1=q[\n]_q+1=[\n+1]_q$.
Introducing 
\begin{equation*}T(A)=f(\n)A+g(\n-1)A^*\textrm{ and }T(A)^*=A^*f(\n)^*+Ag(\n-1)^*\,.\end{equation*}
Consider unitary complex numbers $z,w$ such that
\begin{equation}
zw=1\,.
\end{equation}
For $h(\n)=f(\n)$ or $h(\n)=g(\n)$, assuming
\begin{equation}
h(\n+1)A=zAh(\n)\textrm{ and }Ah(\n)^*=wh(\n+1)^*A \,.
\end{equation} 
For $Q$ a real number we obtain 
\begin{equation*}
[T(A),T(A)^*]_Q=w^2\Phi(\n)A^2+z^2{A^*}^2\Phi(\n)^*+w^2\Psi(\n)\,,
\end{equation*}
where $\Phi(\n)=f(\n)g(\n+1)^*-Qg(\n)^*f(\n+1)$ and
\begin{equation*}
\Psi(\n)=  f(\n)f(\n)^*[\n+1]_q-Qf(\n-1)^*f(\n-1)[\n]_q + g(\n-1)g(\n-1)^*[\n]_q-Qg(\n)^*g(\n)[\n+1]_q\,.
\end{equation*}
To obtain $[T(A),T(A)^*]_Q=1$ we set $\Phi(\n)=0$ and $\Psi(\n)=\frac{1}{w^2}$. 
Therefore, we have $g(\n+1)^*f(\n+1)^{-1}=Qf(\n)^{-1}g(\n)^*$. Set $\zeta:=g(0)^*f(0)^{-1}$ a commuting element with $f(\n)$ and $f(\n)^*$, i.e.,
\begin{equation*}
g(\n)^*f(\n)^{-1}=\zeta Q^{\n}\,.
\end{equation*}
Substituting this relation in the expression $\Psi(\n)=\frac{1}{w^2}$ and defining $F_1(\n)=f(\n)f(\n)^*[\n+1]_q$ and $F_2(\n)=f(\n)^*f(\n)[\n+1]_q$  we obtain 
\begin{equation}
F_1(\n)=\frac{\frac{1}{w^2}+\left(   1-\zeta^*\zeta Q^{2\n-3}   \right)QF_2(\n-1)}{1-\zeta\zeta^*Q^{2\n+1}  }\,.
\end{equation}
Assuming $F(\n):=F_1(\n)=F_2(\n)$, setting $|\zeta|^2:=\zeta\zeta^*=\zeta^*\zeta$ and $\gamma:=F(0)$, this gives a recurrence relation implying the following
\begin{equation}
F(\n)=\frac{1-|\zeta|^2Q^{\n}}{w^2(1-|\zeta|^2Q^{2\n+1})(1-|\zeta|^2Q^{2\n-1})}\left( [\n]_Q+Q^{\n}\Gamma(\n) \right)\,,
\end{equation}
where 
\begin{equation*}
\Gamma(\n)=\gamma w^2\frac{1-|\zeta|^2(Q+Q^{-1})+|\zeta|^4}{1-|\zeta|^2Q^{\n}}\,,
\end{equation*}
and if $f(\n)=f(\n)^*$ we obtain the following value for $f(\n)$ by
\begin{equation}
f(\n)=\sqrt{\frac{1-\zeta^2Q^{\n}}{w^2(1-\zeta^2Q^{2\n+1})(1-\zeta^2Q^{2\n-1})}  \frac{[n+\epsilon_Q(\n)]_Q}{[\n+1]_q} }  \,,
\end{equation}
where 
\begin{equation*}
\epsilon_Q(\n)=\frac{\log(1+(Q-1)\Gamma(\n))}{\log (Q)}\,.
\end{equation*}
For $Q\rightarrow 1$ (and $q=1$) one obtains $\epsilon\rightarrow\gamma w^2(1-\zeta^2)$. The coefficient $f(\n)$ is not constant except for the values $\zeta=0$ and $\gamma=z^2$. When $z=w=1$ the former limit converges to the nonlinear Bogoliubov transformations from \cite{Kat}.

\section{Hopf structure}

An algebra with product and unit, is a Hopf algebra \cite{Majid} when it has a co\-product $\Delta$ and counit $\varepsilon$ and the algebraic anti-homomorphism antipode $\gamma$ sa\-tisfying the following relations 
$$(\Delta\otimes id)\circ\Delta=(id\otimes \Delta)\circ\Delta\,,$$
$$m\circ(\varepsilon\otimes id)\circ\Delta=m\circ(id\otimes\varepsilon)\circ\Delta\,,$$
$$m\circ(\gamma\otimes id)\circ\Delta=m\circ(id\otimes \gamma)\circ\Delta\,.$$
We consider the Hopf structure given in \cite{Hou} for the $q$-deformed Heisenberg algebra, but here we take the generators $A$, $A^*$, $U$, $U^*$, $V$ and $V^*$ 
For $q^d=1$ we assume the identity $A^d=1$ (so $A^{-1}=A^{d-1}$) and if the $*$-operation is preserved, we should have $(A^*)^d=1$, expression for which the physical interpretation is intriguing. In fact, the previous results would imply some cyclic property (condition) when we apply the operators of creation and annihilation to some defined vacuum. For example, consecutive applications of the creation operator over a vacuum $\vert 0\rangle$, $d$ times, would leave the vacuum unchanged. Analogous interpretation applies to the annihilation operator. Given the previous conditions, there exists the following Hopf algebra structure
\begin{itemize}
\item[ ]$\Delta(U)=U\otimes A+A^{-1}\otimes U$,\hspace{1cm}$\Delta(V)=V\otimes A+A^{-1}\otimes V$;
\item[ ]$\Delta(U^*)=U^*\otimes A+A^{-1}\otimes U^*$,\hspace{1cm}$\Delta(V^*)=V^*\otimes A+A^{-1}\otimes V^*$;
\item[ ]$\Delta(A) = A\otimes A$, 
\item[ ]$\Delta(A^*)=A^*\otimes 1+A^{d-2}\otimes A^*+\frac{1}{1-q} \left(A^{d-1}\otimes A^{d-1}\right) \left[1\otimes 1-A^{d-1}\otimes 1-1\otimes A\right]$;
\item[ ]$\epsilon(U)=\varepsilon(U^*)=\varepsilon(V)=\varepsilon(V^*)=0\,,\hspace{1cm}\varepsilon(A)=1\,,$\hspace{1cm}$\varepsilon(A^*)=\frac{1}{1-q}$;
\item[ ]$\gamma(U)=-q^{-1}U$,\hspace{0.1cm} $\gamma(V)=-q^{-1}V$,\hspace{0.1cm} $\gamma(U^*)=-qU^*$,\hspace{0.1cm} $\gamma(V^*)=-qV^*$,
\item[ ]$\gamma(A)=A^{d-1}$,\hspace{0.5cm}$\gamma(A^*)=-A^2A^*+\frac{2A}{1-q}$.
\end{itemize}
In order to obtain  $\Delta$ and $\gamma$ for the product of two elements we use that $\gamma$ is an antialgebra map and the bialgebra property 
$$\gamma\circ m=m\circ\tau\circ(\gamma\otimes\gamma)\,,$$
$$\Delta\circ m=(m\otimes m)\circ(id\otimes \tau\otimes id)\circ(\Delta\otimes\Delta)\,,$$
where $\tau$ is the twist linear map. \\

\section*{Acknowledgements}
\noindent I. A. is supported by the JSPS Postdoctoral fellow for oversea researchers. \\ C. S. is supported by C\'atedras CONACYT.


\section*{References}

\begin{thebibliography}{99}

\bibitem{Bog0} N. N. Bogoliubov, On a New Method in the Theory of Superconductivity. Il Nuevo Cimento, 1958, Vol. 7 N. 6, 794-805. 



 
\bibitem{Hou} Bo-Yuan Hou and Lian-Chao Xu: The Hopf Algebraic Structure of $q$-Deformed Heisenberg Algebra when $q$ Is a Root of Unity. Commun. Theor. Phys., 1995, 24, 481-482.





\bibitem{Kat} J. Katriel: A nonlinear Bogoliubov Transformation. Physics Letters A, 2003, 307, 1-7. 


\bibitem{Majid} S. Majid: A Quantum Groups Primer. Cambridge University Press, 2002.

\bibitem{Man} Yu. I. Manin: Quantum groups and noncommutative geometry. Universit\'e de Montr\'eal, Centre de Recherches Math\'ematiques, Montreal, QC, 1988. 




\bibitem{Parker} L.E. Parker and D. J. Toms: Quantum Field Theory in Curved Spacetime. Cambridge University Press, 2009.

\bibitem{Val}  J. G. Valatin: Comments on the theory of superconductivity. Il Nuevo Cimento, 1958, Vol. 7, N. 6, 843-857.

\bibitem{Xin} Xin Zhang: Black hole evaporation based upon a q-deformation description, Internat. J. Modern Phys, 20 (2005) No. 26, 6039-6049.

\bibitem{Zhe} A. S. Zhedanov: Bogoliubov $q$-transformations and Clebsch-Gordan coefficients for a $q$-oscillator. Physics Letters A, 2003, 307, 1-7. 



\end{thebibliography}
\end{document}